\documentclass[%
reprint,
 amsmath,amssymb,
 aps,
]{revtex4-2}
\usepackage[letterpaper,top=2cm,bottom=2cm,left=3cm,right=3cm,marginparwidth=1.75cm]{geometry}
\usepackage{graphicx}
\usepackage{dcolumn}
\usepackage{amsmath}
\usepackage[bottom]{footmisc}
\usepackage{bm}

\begin{document}


\title{Natural Emergence of $\Lambda$CDM Cosmology within General Relativity from Two Alternative Frameworks Without Fine-Tuning and Coincidence}
\author{H. R. Fazlollahi}
\email{shr.fazlollahi@gmail.com (Corresponding Author)}
\affiliation{%
 PPGCOSMO \& Departamento de Física, Universidade Federal do Espirito Santo (UFES), Av. Fernando Ferrari, 514 Campus de Goiabeiras, Vitória, Espírito Santo CEP 29075-910, Brazil}%

\begin{abstract}
In this study, by revisiting the quantum interpretation of the cosmological constant, we introduce its formal representation within standard General Relativity. Examining its behavior in a Friedmann–Robertson–Walker spacetime reveals a mechanism in which the symmetry between energy and momentum is dynamically broken. Applying this concept naturally leads to the derivation of the familiar $\Lambda$CDM model, while simultaneously alleviating both the fine-tuning and coincidence problems. Comparison of the ground-state energy behavior in the Friedmann equations with a dust matter field further indicates that large-scale matter exhibits the same symmetry-breaking behavior. Remarkably, due to this broken symmetry, the interactions between local regions of matter in the large-scale structure generate effective pressure, driving late-time acceleration and reproducing the $\Lambda$CDM expansion history without invoking exotic fields or negative-pressure components. This framework provides a self-consistent realization of $\Lambda$CDM within General Relativity, emerging entirely from the intrinsic dynamics of standard matter without fine-tuning and coincidence problems.
\begin{description}
\item[Keywords]
Late-Time Cosmic Acceleration; Energy–Momentum Symmetry Breaking; Matter Self-Interactions; Cosmological Constant Problems.
\end{description}
\end{abstract}

\maketitle

\section{Introduction}

The discovery that the Universe is undergoing accelerated expansion represents one of the most profound challenges in modern cosmology. Initially revealed by high-redshift Type Ia supernova observations~\cite{SupernovaSearchTeam:1998fmf, SupernovaCosmologyProject:1998vns}, this acceleration has since been confirmed through complementary probes, including baryon acoustic oscillations (BAO)~\cite{BOSS:2012dmf, SDSS:2005xqv}, measurements of cosmic microwave background (CMB) anisotropies~\cite{Planck:2018vyg, WMAP:2003ivt}. Collectively, these observations indicate that roughly 70\% of the current energy density of the Universe is associated with a component driving accelerated expansion, commonly referred to as dark energy. Despite this compelling empirical evidence, the physical origin and fundamental nature of this component remain undetermined, motivating the exploration of a wide range of theoretical models.

Among the proposed explanations, the simplest and historically earliest approach consists of introducing a cosmological constant $\Lambda$ into the Einstein–Hilbert action,
\begin{equation}
S = \int \sqrt{-g}\,(R - 2\Lambda)\, d^4x + S_m ,
\end{equation}
which leads to the modified Einstein field equations
\begin{equation}
G_{\mu\nu} + \Lambda g_{\mu\nu} = T_{\mu\nu}.
\end{equation}
This construction gives rise to the $\Lambda$CDM model, which has achieved remarkable empirical success. It accurately reproduces the expansion history of a homogeneous and isotropic Universe across radiation, matter, and late-time epochs~\cite{SupernovaCosmologyProject:1998vns, Planck:2018vyg}. Moreover, it provides a consistent framework for interpreting CMB anisotropies, large-scale structure formation, and gravitational lensing observations, while involving a minimal set of free parameters. These features have established $\Lambda$CDM as the standard model of cosmology~\cite{Carroll:2000fy}.

Despite its phenomenological success, the $\Lambda$CDM framework faces significant theoretical challenges. Foremost among them is the cosmological constant problem: identifying $\Lambda$ with the vacuum energy predicted by quantum field theory yields theoretical estimates that exceed the observed value by more than 120 orders of magnitude~\cite{Padmanabhan:2002ji, Weinberg:1988cp}, constituting an extreme fine-tuning problem. Closely related is the coincidence problem, which questions why the energy densities of matter and dark energy are of the same order precisely in the present cosmological epoch~\cite{Carroll:2000fy, Weinberg:1988cp}. These conceptual difficulties suggest that, although observationally successful, $\Lambda$CDM may not represent a complete fundamental description.

In addition to these theoretical concerns, several observational tensions have emerged. The so-called Hubble tension, the discrepancy between the locally measured value of the Hubble constant, $H_0\sim 73\, \mathrm{km\, s^{-1}\, Mpc^{-1}}$~\cite{SupernovaCosmologyProject:1998vns, Planck:2018vyg}, and the value inferred from CMB measurements within $\Lambda$CDM, $H_0 \sim 67\, \mathrm{km\, s^{-1}\, Mpc^{-1}}$~\cite{Riess:2021jrx, Riess:2016jrr}, may point to new physics or unresolved systematic effects. Furthermore, mild discrepancies in the amplitude of matter fluctuations, quantified by $\sigma_8$, as inferred from weak lensing and galaxy surveys~\cite{DiValentino:2020vvd, Nunes:2021ipq}, suggest possible deviations from the standard scenario. While mechanisms such as primordial magnetic fields at recombination~\cite{Jedamzik:2020krr, Jedamzik:2025cax} can alleviate certain tensions, the fundamental fine-tuning and coincidence problems remain unresolved within the conventional $\Lambda$CDM paradigm.

These issues have motivated extensive exploration of modified gravity frameworks, including $f(R)$ theories~\cite{Sotiriou:2008rp}, scalar–tensor theories~\cite{Capozziello:2011et}, and emergent gravity scenarios~\cite{Rastall:1976uh, Fazlollahi:2023rhg}. Although such models may address specific conceptual challenges, they often introduce additional degrees of freedom, enlarged parameter spaces, or complex dynamics that are not tightly constrained by observations. Moreover, many encounter difficulties in reproducing the matter-dominated expansion history with the same level of consistency achieved by $\Lambda$CDM~\cite{Tsujikawa:2007gd, Berti:2015itd}. Consequently, despite its theoretical shortcomings, $\Lambda$CDM remains unparalleled in its combination of simplicity, predictive power, and observational agreement.

In this work, motivated by the concept of ground-state energy in quantum field theory, we begin in Sec.~II by re-examining the $\Lambda$CDM framework from a foundational perspective, demonstrating that the geometrical cosmological constant $\Lambda$ is not formally consistent with the ground-state energy interpretation commonly adopted in quantum field theory. In Secs.~III and IV, we introduce the notion of energy--momentum symmetry breaking associated with the ground-state sector and formulate a mathematically consistent mechanism that naturally addresses the coincidence problem while substantially alleviating the fine-tuning tension inherent in the standard $\Lambda$CDM paradigm. In Sec.~V, we show that the analogous behavior between ground-state energy and ordinary matter in breaking the symmetry between energy and momentum motivates extending this mechanism beyond the vacuum sector. Within this generalized framework, matter itself, through the same symmetry-breaking structure, gives rise to late-time cosmic acceleration without invoking an explicit dark energy component. Importantly, this construction reproduces the $\Lambda$CDM expansion history while avoiding its fine-tuning and coincidence problems. Finally, Sec.~VI summarizes the principal results and discusses their broader theoretical and phenomenological implications.
\par\vspace{2em}
Throughout this work, we adopt units in which the reduced Planck mass is set to unity, $m_P^2 = \kappa^{-1} = 1$, and all $c_i$ are regarded as integration constants of the model.

\section{Ground-State Energy in the $\Lambda$ Paradigm}

Symmetry principles provide the foundational constraints governing the formulation of fundamental physics \cite{Weinberg:1995mt, Weinberg:1980wa}. In relativistic field theory, invariance under spacetime translations implies conservation of the energy–momentum four-vector, while Lorentz symmetry unifies energy and spatial momentum into a single geometric entity \cite{Misner:1973prb}. Despite this unification at the level of representation theory, the structure of the quantum vacuum introduces a nontrivial distinction between these components.

In quantum field theory, the ground state $|0\rangle$ satisfies \cite{Weinberg:1995mt, Peskin:1995ev}
\begin{equation}
\mathcal{E}_0 = \langle 0 | H | 0 \rangle ,
\end{equation}
\begin{equation}
P^\mu |0\rangle = 0 .
\end{equation}
Although zero-point fluctuations contribute $\frac{1}{2}\hbar \omega_k$ per mode to the Hamiltonian, translational invariance and spatial isotropy ensure that the expectation value of the momentum operator vanishes identically. The vacuum therefore, possesses a nonvanishing energy density while carrying no net spatial momentum. Ground-state energy is intrinsically associated with the temporal sector and does not induce momentum flux.

In gravitational theory, the cosmological constant $\Lambda$ is commonly interpreted as the macroscopic manifestation of vacuum (ground-state) energy. Within the Einstein–Hilbert framework, it enters geometrically through Eq. (2).
For a spatially flat Friedmann–Robertson–Walker (FRW) spacetime filled with pressureless matter, the corresponding Friedmann equations read \cite{Carroll:2000fy}
\begin{equation}
3H^2 = \rho_m + \Lambda ,
\end{equation}
\begin{equation}
2\dot{H} + 3H^2 = \Lambda ,
\end{equation}
which correspond to an effective perfect fluid with equation of state
\begin{equation}
p_\Lambda = -\rho_\Lambda = -\Lambda .
\end{equation}
In this formulation, the $\Lambda$ term contributes symmetrically to both temporal and spatial components of the gravitational field equations, reflecting its proportionality to $g_{\mu\nu}$. A structural tension then becomes evident. While the quantum vacuum carries energy without momentum, the geometric $\Lambda$ term enforces maximal spacetime symmetry by contributing equally to all components of the metric tensor. Consequently, the conventional identification of $\Lambda$ with ground-state energy does not faithfully reproduce the defining vacuum property $P^\mu = 0$ at the level of the gravitational source.

If $\Lambda$ is fundamentally associated with ground-state energy, its gravitational implementation should reflect its purely energy-like character. A covariant realization of this idea can be achieved by introducing a normalized timelike vector field $u_\mu$, satisfying
\begin{equation}
u_\mu u^\mu = \pm 1 ,
\end{equation}
where the sign corresponds to the chosen metric signature convention: for $(+,-,-,-)$, $u_\mu u^\mu = +1$, while for $(-,+,+,+)$, $u_\mu u^\mu = -1$. This normalization ensures that $u^\mu$ defines a unit timelike direction, naturally interpreted as the four-velocity of the comoving frame.

We then consider modified gravitational field equations of the form \cite{Fazlollahi:2026kbv}
\begin{equation}
G_{\mu\nu} = T_{\mu\nu} \pm \Lambda u_\mu u_\nu ,
\end{equation}
where the overall sign is chosen consistently with the normalization condition to guarantee that the energy contribution is positive in the comoving frame. In a spatially flat FRW background with $u^\mu = (1,0,0,0)$ in comoving coordinates, the corresponding Friedmann equations read
\begin{equation}
3 H^2 = \rho_m + \Lambda ,
\end{equation}
\begin{equation}
2 \dot{H} + 3 H^2 = 0 .
\end{equation}
In this formulation, $\Lambda$ contributes exclusively to the energy density equation while leaving the acceleration equation unaltered. Consequently, the additional term behaves as a purely timelike energy component, consistent with the defining vacuum (ground-state) energy condition $P^\mu = 0$.

These field equations can be derived from the covariant action \cite{Fazlollahi:2026kbv}
\begin{equation}
S = \int d^4 x \, \sqrt{-g} \left[ \frac{R}{2} + \mathcal{L}_m - \frac{\Lambda}{2}\left( g^{\mu\nu} u_\mu u_\nu \mp 1 \right) \right] ,
\end{equation}
where the constraint term enforces $u_\mu u^\mu = \pm 1$ through variation with respect to $\Lambda$. The choice of sign matches the metric signature and ensures internal consistency between the action and the resulting field equations.

The structural distinction between this construction and the conventional $\Lambda$-term is clear. While the standard cosmological constant enforces complete spacetime symmetry in the stress-energy tensor, the present formulation singles out the temporal direction preferred by the vacuum state. The resulting source preserves covariance but does not impose isotropic pressure. 

This analysis demonstrates that the conventional geometric implementation of $\Lambda$, proportional to $g_{\mu\nu}$, does not faithfully reflect the defining structural property of ground-state energy in quantum field theory. By contrast, the action above and its associated field equations provide a covariant realization in which the vacuum contribution retains its intrinsically timelike character. Within this framework, $\Lambda$ is consistently interpreted as a ground-state energy component embedded in gravitational dynamics. Consequently, if $\Lambda$ is interpreted as the macroscopic manifestation of vacuum (ground-state) energy, the corresponding gravitational field equations and action are naturally given by Eqs.~(9) and (12), respectively.

\section{Energy–Momentum Symmetry and Its Breaking in Gravitational Dynamics}

The interpretation of $\Lambda$ as ground-state energy in Sec.~II leads naturally to the action (12) and the corresponding field equations (9). For a spatially flat FRW spacetime filled with pressureless matter, the resulting Friedmann equations (10) and (11) exhibit a distinctive feature: the vacuum contribution appears in the first Friedmann equation, while the second remains unmodified. Accordingly, no explicit pressure term is associated with $\Lambda$ at the level of the field equations. This observation raises a fundamental dynamical question: how can a macroscopic energy component without an associated pressure drive late-time cosmic acceleration?

To clarify this point, it is instructive to compare our framework with the general structure of modified gravitational dynamics. In a broad class of alternative theories of gravity, the field equations may be written as
\begin{equation}
    G_{\mu\nu} = T_{\mu\nu} + \mathcal{M}_{\mu\nu},
\end{equation}
where $\mathcal{M}_{\mu\nu}$ encodes additional geometric or effective contributions beyond standard General Relativity. In this sense, $\mathcal{M}_{\mu\nu}$ represents the full modification sector of alternative gravity theories.

For a spatially flat FRW background with dust matter, these equations reduce to
\begin{equation}
    3H^2=\rho_m+\mathcal{M}_{tt},
\end{equation}
\begin{equation}
    2\dot{H}+3H^2=-\mathcal{M}_{ii},
\end{equation}
where $\mathcal{M}_{tt}$ and $\mathcal{M}_{ii}$ denote the temporal and spatial components of the modification sector, respectively.

In generic modified gravity scenarios, temporal and spatial corrections appear simultaneously: a nonvanishing $\mathcal{M}_{tt}$ is typically accompanied by a corresponding spatial component $\mathcal{M}_{ii}$. The symmetry between the energy (temporal) and momentum/pressure (spatial) sectors is therefore preserved at the level of the effective source.

By contrast, the structure derived in Sec.~II corresponds to
\begin{equation}
    3H^2=\rho_m+\mathcal{M}_{tt},
\end{equation}
\begin{equation}
    2\dot{H}+3H^2=0,
\end{equation}
with $\mathcal{M}_{ii}=0$. Here, the correction affects the temporal component without introducing a spatial counterpart. The symmetry between energy and momentum sectors within the modification term is thus broken: additional energy density is present, yet no direct pressure contribution appears in the field equations. This feature directly reflects the vacuum property $P^\mu = 0$.

The physical implications of this symmetry breaking become evident at the level of the continuity equation \cite{Fazlollahi:2026kbv, Fazlollahi:2024hud}. Taking the covariant divergence of the modified field equations yields
\begin{equation}
    \rho_m' + \mathcal{M}_{tt}' + 3 (\rho_m + \mathcal{M}_{tt}) = 0 ,
\end{equation}
where a prime denotes differentiation with respect to $x = \ln a$. Because the modification lacks a spatial component, the continuity equation naturally accommodates two consistent interpretations.

\subsection*{A. Effective Single-Component Interpretation}
Defining the total energy density as
\begin{equation}
    \rho_{\rm tot} =\rho_m+\mathcal{M}_{tt},
\end{equation}
the continuity equation reduces to
\begin{equation}
    \rho_{\rm tot}'+3\rho_{\rm tot}=0 ,
\end{equation}
which admits the straightforward solution
\begin{equation}
    \rho_{\rm tot}\propto e^{-3x} .
\end{equation}

\subsection*{B. Effective Pressure Interpretation}

Alternatively, the continuity equation can be rearranged as
\begin{equation}
    \rho_m'+3\Big(\rho_m+\mathcal{M}_{tt}+\frac{1}{3} \mathcal{M}_{tt}'\Big) = 0 .
\end{equation}
This motivates the definition of an effective matter component $\tilde{\rho}_m$ with pressure
\begin{equation}
    \tilde{p}=\mathcal{M}_{tt}+\frac{1}{3}\mathcal{M}_{tt}' ,
\end{equation}
Such that the continuity equation assumes the standard form
\begin{equation}
    \tilde{\rho}_m'+3(\tilde{\rho}_m+\tilde{p})=0 ,
\end{equation}
corresponding to the effective Friedmann system
\begin{equation}
    3 H^2=\tilde{\rho}_m,
\end{equation}
\begin{equation}
    2\dot{H}+3H^2=-\tilde{p}.
\end{equation}
Although the modification enters the gravitational equations exclusively through the temporal component, it dynamically induces an effective pressure via the evolution equation. The pressure is therefore not introduced at the level of the field equations but emerges naturally from the broken energy–momentum structure \cite{Fazlollahi:2026kbv}.

At this stage, a crucial conceptual question arises: by defining an effective pressure from the temporal modification, are we merely performing a formal rearrangement, or does this procedure correspond to the emergence of a genuinely new matter sector? The answer lies in the dynamical content of the conservation law. Once the modification term appears as the pressure, the original matter component can no longer evolve independently as pressureless dust. Interpreting the additional contributions as pressure necessarily modifies the evolution of the matter sector. The effective fluid obeying a standard conservation equation with nonvanishing pressure is thus dynamically distinct from the original dust component.

This distinction becomes transparent when the total system is expressed as two interacting components \cite{Fazlollahi:2026kbv}:
\begin{equation}
    \rho_m'+3\rho_m=\mathcal{Q},
\end{equation}
\begin{equation}
    \mathcal{M}_{tt}'+3\mathcal{M}_{tt}=-\mathcal{Q} ,
\end{equation}
where $\mathcal{Q}$ characterizes the energy exchange between the matter sector and the modification sector. Comparing the second equation with the definition of the effective pressure immediately gives
\begin{equation}
    \tilde{p} = -\frac{\mathcal{Q}}{3} .
\end{equation}
This relation clarifies the situation: the effective pressure is the dynamical manifestation of the interaction term in the decoupled picture. Whenever $\mathcal{Q}\neq 0$, the matter component does not follow standard dust evolution, and the system must be described as an interacting fluid with nonvanishing pressure. Hence, defining the effective pressure is not a mere change of variables. It reflects the fact that, once the symmetry between the temporal and spatial sectors is broken, energy exchange between components becomes unavoidable at the level of the continuity equation. The resulting fluid is dynamically distinct from ordinary dust, and its effective pressure provides the mechanism by which a ground-state energy contribution can drive late-time cosmic acceleration.

\section{Dynamical $\Lambda$ from Symmetry Breaking}

To assess whether the energy–momentum symmetry-breaking formalism indeed yields a late-time accelerating phase, we reconsider the modified Friedmann equations (10) and (11) within the present framework. Here, the temporal component of the modification tensor is identified with the cosmological constant as the macroscopic manifestation of ground-state energy:
\begin{equation}
    \mathcal{M}_{tt}=\Lambda=\zeta\mathcal{E}_0 ,\quad\quad\mathcal{M}_{ii}=0,
\end{equation}
where $\mathcal{E}_0$ denotes the ground-state energy density arising from the underlying quantum medium, and $\zeta$ is a proportionality constant connecting microscopic (quantum) physics to macroscopic (cosmological) dynamics.

Using the decoupled conservation equation, one immediately obtains
\begin{equation}
    3\zeta\mathcal{E}_0=-\mathcal{Q},
\end{equation}
implying that the interaction term $\mathcal{Q}$ is nonvanishing. Consequently, an effective pressure can be defined as
\begin{equation}
    \tilde{p} =-\frac{\mathcal{Q}}{3}=\zeta \mathcal{E}_0 .
\end{equation}
The corresponding continuity equation then takes the standard form
\begin{equation}
    \tilde{\rho}_m'+3(\tilde{\rho}_m+\zeta \mathcal{E}_0)=0,
\end{equation}
equivalently described by the modified Friedmann system
\begin{equation}
    3H^2=\tilde{\rho}_m,
\end{equation}
\begin{equation}
    2\dot{H}+3H^2=-\zeta\mathcal{E}_0.
\end{equation}
Solving these equations (or equivalently the continuity equation) yields
\begin{equation}
    \tilde{\rho}_m=\rho_{m0}e^{-3x} - \zeta\mathcal{E}_0 ,
\end{equation}
so that the Friedmann equations can be written as
\begin{equation}
    3H^2=\rho_{m0} e^{-3x}-\zeta\mathcal{E}_0,
\end{equation}
\begin{equation}
    2\dot{H}+3H^2=-\zeta\mathcal{E}_0.
\end{equation}
These expressions are formally identical to those of the standard $\Lambda$CDM scenario, with the correspondence \cite{Fazlollahi:2026kbv}
\begin{equation}
    \rho_\Lambda =-p_\Lambda=-\zeta\mathcal{E}_0.
\end{equation}
However, the conceptual origin is fundamentally different. In the present construction, the effective cosmological constant emerges from energy–momentum symmetry breaking applied to the ground-state energy. Furthermore, the interaction term $\mathcal{Q}$ is not introduced ad hoc, but emerges directly from the intrinsic nature of the ground-state energy. The formalism therefore predicts an intrinsic coupling between dust matter and ground-state energy, establishing a dynamical relation between them. This feature has immediate implications for the coincidence problem: since matter and ground-state energy are dynamically connected through $\mathcal{Q}$, their relative magnitudes are no longer independent, providing a natural mechanism linking their present-day energy densities.

Regarding the fine-tuning problem, the proportionality constant $\zeta$ plays a central role. From the relation
\begin{equation}
    \frac{\Omega_{\mathcal{E}_0}}{\Omega_\Lambda}=- \frac{1}{\zeta},
\end{equation}
and using the observed hierarchy of energy scales, one obtains \cite{Padmanabhan:2002ji, Weinberg:1988cp}
\begin{equation}
    \zeta\sim- 10^{-123} .
\end{equation}
Remarkably, this magnitude can be expressed in terms of fundamental constants as \cite{Fazlollahi:2026kbv}
\begin{equation}
    |\zeta|\sim \frac{\hbar G H_0^2}{c^5} ,
\end{equation}
demonstrating that $\zeta$ encodes a precise combination of quantum ($\hbar$), gravitational ($G$), and cosmological ($H_0$) scales. In this sense, it explicitly bridges the microscopic ground-state energy $\mathcal{E}_0$ and its macroscopic gravitational manifestation $\Lambda$.

The effective-pressure interpretation thus demonstrates that the ground-state energy density is not gravitationally inert, but dynamically couples to the dust matter sector through the symmetry-breaking structure of the field equations. This coupling inevitably generates negative pressure, driving a late-time accelerating phase. While the resulting background evolution reproduces the expansion history of $\Lambda$CDM, the physical origin is fundamentally distinct: cosmic acceleration emerges as a dynamical consequence of energy–momentum symmetry breaking in the vacuum sector, rather than from postulating a purely geometric cosmological constant at the level of the action.

\section{Cosmos Without a Dark Energy Sector}

A reexamination of the ground-state energy within the present framework reveals a noteworthy structural feature: although it may be formally recast as an effective dark energy contribution, its dynamical role is indistinguishable from that of pressureless dust. In both cases, no spatial pressure component is present. For the ground-state energy, this follows from its intrinsic definition in quantum field theory; for dust matter, it is imposed by construction. At the level of the Friedmann equations, each contributes exclusively through its temporal energy density.

This structural correspondence indicates that late-time acceleration need not originate from an independent dark energy sector. If a pressureless ground-state contribution acquires cosmological significance through symmetry breaking in the energy--momentum structure, it becomes natural to investigate whether the dust matter sector itself may exhibit an analogous dynamical reorganization. Within such a scenario, the onset of acceleration is traced back to the internal dynamics of the matter sector rather than to an additional exotic component.

To analyze this possibility, the ground-state contribution is set aside, reducing the system to the standard matter-dominated Friedmann equations,
\begin{equation}
    3H^2=\rho_m,
\end{equation}
\begin{equation}
    2\dot H+3H^2=0.
\end{equation}
Under the conventional interpretation, accelerated expansion is excluded because pressureless dust carries no intrinsic momentum flux, and the second Friedmann equation reflects vanishing pressure. However, explicit implementation of energy–momentum symmetry breaking permits the temporal energy density to be written as a linear combination of two dynamically distinguishable contributions.
\begin{equation}
3H^2 =\kappa\left(\alpha\rho_m+\beta\rho_m\right),
\qquad\alpha+\beta = 1 .
\end{equation}
This decomposition does not introduce additional degrees of freedom. Instead, it reveals a two-sector dynamical structure inherent in the matter field. One sector retains the behavior of conventional matter field, while the remaining sector effectively plays the role of an auxiliary contribution analogous to $\mathcal{M}_{tt}$.

The continuity equation correspondingly acquires the composite form
\begin{equation}
    \alpha\rho_m'+\beta\rho_m'+3(\alpha\rho_m +\beta\rho_m)=0.
\end{equation}
Decoupling this relation yields
\begin{equation}
    \alpha\rho_m'+3\alpha\rho_m=\mathcal{Q},
\end{equation}
\begin{equation}
    \beta\rho_m'+3\beta\rho_m=-\mathcal{Q},
\end{equation}
The interaction term arises as a direct consequence of the decomposition and does not represent an external coupling.

Alternatively, the $\beta$-sector contributions may be absorbed into an effective pressure, in which case the continuity equation takes the form 
\begin{equation}
    \alpha\rho_m'+3\left(\alpha\rho_m+\beta\rho_m+\frac{\beta}{3}\rho_m'\right)=0,
\end{equation}
from which the emergent pressure is identified as
\begin{equation}
    \tilde p =\beta\rho_m+\frac{\beta}{3}\rho_m'.
\end{equation}
For $\mathcal{Q}\neq 0$, the redefined fluid satisfies the standard conservation law
\begin{equation}
    \tilde\rho'+3(\tilde\rho+\tilde p)=0,
\end{equation}
with the $\alpha$-sector absorbed into the effective density and the $\beta$-sector generating the pressure contribution.

Given that the system contains a single matter field, the interaction may be parametrized as
\begin{equation}
    \mathcal{Q}=3\gamma\rho_m .
\end{equation}
Substitution into the decoupled relations obtains
\begin{equation}
\mathcal{Q}=3\gamma c_1 e^{-\sigma x}, \qquad 
\sigma=\frac{3(\beta+\gamma)}{\beta}.
\end{equation}
The resulting interaction originates from internal energy exchange between the two matter sectors and leads to the following global effective pressure and density: 
\begin{equation}
    \tilde p=-c_1\gamma e^{-\sigma x},
\end{equation}
\begin{equation}
    \tilde\rho=-c_1\beta e^{-\sigma x}+c_2 e^{-3x}.
\end{equation}
In the present case, the same formal procedure is followed here. The difference is that, in the present case, the interaction term is introduced phenomenologically, whereas in the previous section it emerged directly from the underlying structure of the model. The dynamical treatment, however, remains identical.

The matter field in large-scale structure behaves as dust, reproducing the standard isotropic and homogeneous evolution at cosmological scales. Yet, when the internal dynamics implied by energy–momentum symmetry breaking are taken into account, this simple picture no longer holds. Local interactions between the constituent sectors of the matter field generate an emergent effective pressure and alter the evolution of the energy density. In this framework, late-time acceleration arises from a redistribution of energy within the matter sector itself. No independent dark energy component is required; the accelerated phase emerges from the intrinsic dynamical structure of matter once symmetry between temporal and spatial sectors is broken.

Before confronting the model with observational data, it is crucial to verify its internal consistency. The construction introduces two parameters, $\alpha$ and $\beta$, through the indirect interpretation of the continuity equation. This decomposition, however, is inherently ambiguous: one could equivalently assign either the $\alpha$-dependent or the $\beta$-dependent sector to the effective pressure without altering the formal structure of the equations. Physical predictions must remain invariant under such a bookkeeping choice.

Imposing invariance under this redefinition enforces a strict symmetry between the two sectors. The unique consistent solution is
\begin{equation}
    \alpha=\beta=1/2,
\end{equation}
ensuring that the effective theory is fully symmetric and independent of which sector is designated as pressure or as density. This symmetry not only resolves the decomposition ambiguity but also reinforces the interpretation of the effective pressure as an emergent property arising intrinsically from the internal dynamics of the dust matter field.

Adopting the symmetric choice $\alpha=\beta=0.5$ allows for a natural simplification of the model. In its minimal realization, by setting $\gamma=-\beta=-0.5$, the effective density and pressure take the form
\begin{equation}
    \tilde{\rho}=-\frac{c_1}{2}+\rho_{m0} e^{-3x},
\end{equation}
\begin{equation}
    \tilde{p}=\frac{c_1}{2},
\end{equation}
with $c_2 = \rho_{m0}$. In this simplest realization, the model formally reproduces a $\Lambda$CDM–like expansion; however, the underlying mechanism is fundamentally distinct. No additional component is introduced: the dynamics arise entirely from the dust matter field and the collective effect of intrinsic interactions among its local constituents. Within this framework, both the effective pressure $\tilde{p}$ and the modified density $\tilde{\rho}$ emerge from the global aggregation of these internal interactions, demonstrating that ordinary dust matter alone can generate the large-scale behavior traditionally attributed to a cosmological constant.

This minimal model also offers conceptual advantages relative to the ground-state energy formalism discussed in previous sections. In the ground-state energy scenario, an extra field is explicitly introduced to break energy–momentum symmetry, and the combination of the interaction term and a free proportionality parameter addresses both the fine-tuning and coincidence problems. In contrast, the dust-only realization requires no additional fields, and the standard issues associated with the cosmological constant, fine-tuning and coincidence, simply do not arise. Even in this minimal form, the emergent cosmological behavior reproduces the $\Lambda$CDM background evolution without invoking exotic components or quantum vacuum effects.

For arbitrary values of the interaction parameter $\gamma$, the model can be confronted with observational data to test its consistency with the measured expansion history and large-scale structure. Imposing the boundary condition $\omega_{\rm tot} = \omega_0$ and the normalization $H^2(x=0)/H_0^2=1$ at the present epoch determines
\begin{equation}
    c_2=3H_0^2+\frac{c_1}{2},
\end{equation}
\begin{equation}
    \omega_0=-\frac{\gamma c_1}{3 H_0^2}.
\end{equation}
Further constraints can be obtained from the deceleration parameter
\begin{equation}
    q=\frac{1}{2}\left(1+\frac{3\tilde{p}}{\tilde{\rho}}\right),
\end{equation}
which must vanish at the transition redshift $z_T$ \cite{Fazlollahi:2022kbv}, yielding
\begin{equation}
    c_1=\frac{6H_0^2}{(6\gamma-1)(1+z_T)^{6\gamma}}.
\end{equation}
Consequently, for general $\gamma$, the theory depends on the three cosmological parameters $\{H_0, z_T, \gamma\}$. Their best-fit values are obtained via a Markov Chain Monte Carlo (MCMC) analysis of the Pantheon+SH0ES supernova dataset \cite{Hastings:1970aa, Foreman-Mackey:2012any}. The parameter estimates and marginalized posterior distributions are given by following table 

\begin{table}[t]
\centering
\renewcommand{\arraystretch}{1.25}
\begin{tabular}{lc}
\hline\hline
Parameter & Best-fit value (68\% C.L.) \\
\hline
$\quad\quad H_0$ & $67.40^{+0.01}_{-0.01}$ \\[4pt]
$\quad\quad z_T$ & $0.639^{+0.009}_{-0.009}$ \\[4pt]
$\quad\quad \gamma$ & $-0.500^{+0.010}_{-0.011}$ \\
\hline\hline
\end{tabular}
\caption{Best-fit values of the model parameters obtained from the MCMC analysis of the Pantheon+SH0ES compilation.}
\label{tab:params}
\renewcommand{\arraystretch}{1}
\end{table}

\begin{figure}[h!]
\centering
\includegraphics[width=0.48\textwidth]{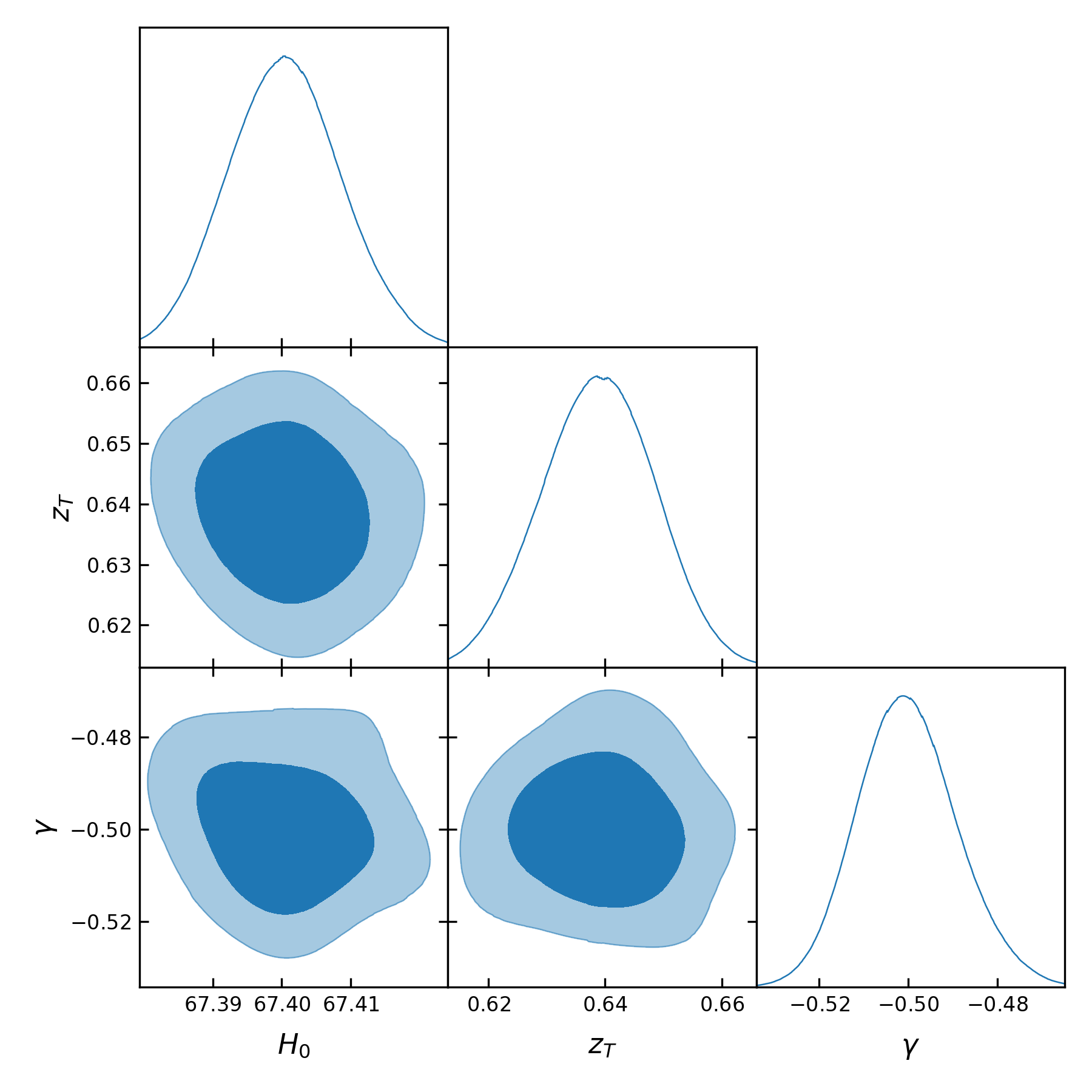}
\caption{Marginalized posterior distributions and confidence regions for the parameter set $\{H_0, z_T, \gamma\}$ from the MCMC analysis of the Pantheon+SH0ES dataset.}
\label{fig:MCMC}
\end{figure}

\begin{figure}[h!]
\centering
\includegraphics[width=0.48\textwidth]{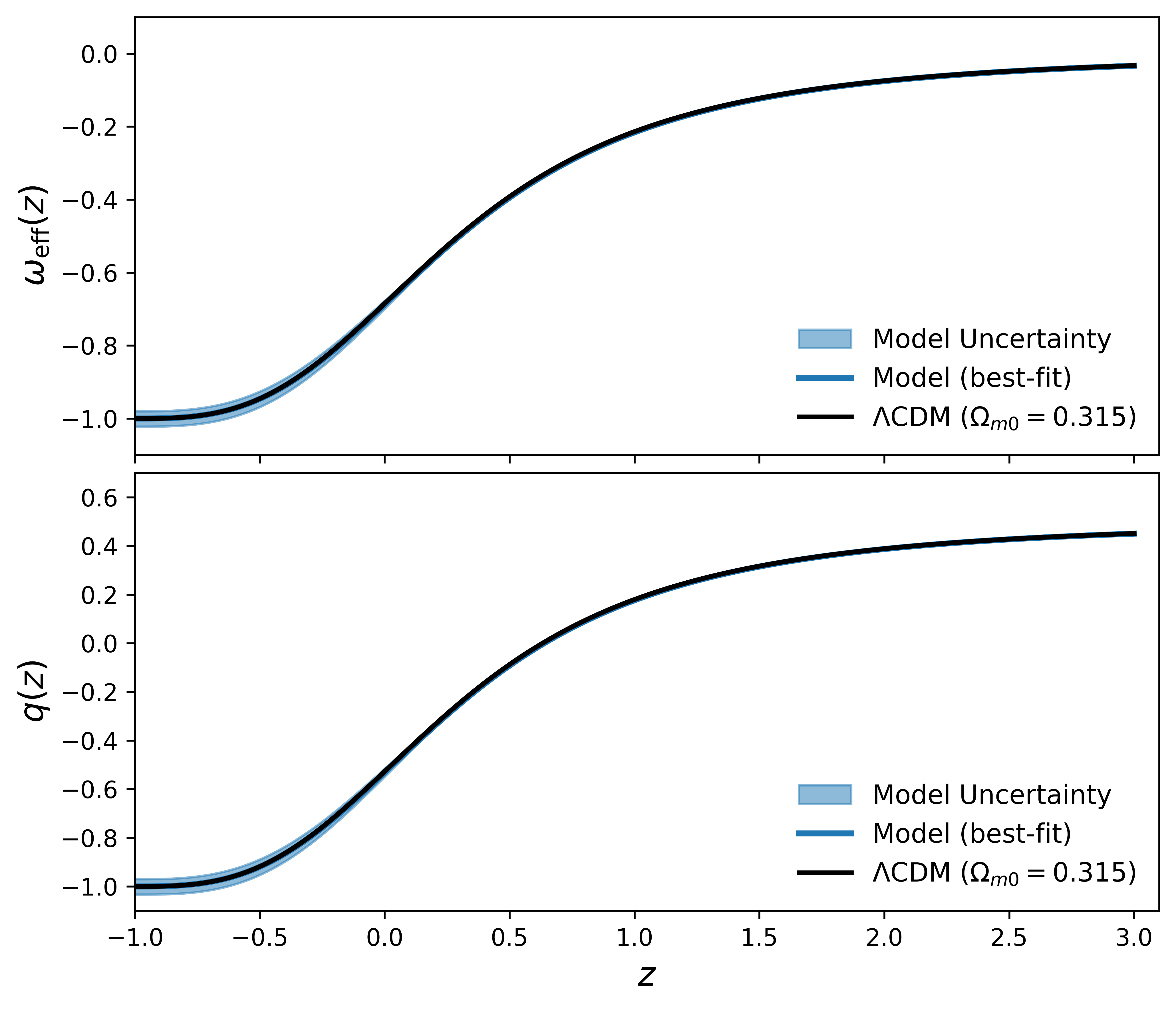}
\caption{Redshift evolution of the total equation-of-state parameter $\omega_{\rm tot}$ (top) and deceleration parameter $q$ (bottom) compared with the $\Lambda$CDM model.}
\label{fig:evolution}
\end{figure}

Comparison with Planck 2018 results demonstrates that the minimal self-matter interaction model is remarkably consistent with the standard cosmological constant scenario \cite{Planck:2018vyg}. In particular, the median value $\gamma \simeq -0.5$ ensures that the background dynamics closely reproduce $\Lambda$CDM. As a result, both the ground-state energy scenario and the minimal dust-only framework produce convergent predictions, capturing the late-time acceleration and the $\Lambda$CDM expansion history (Figure~\ref{fig:evolution}).

This analysis establishes that the $\Lambda$CDM expansion history can be obtained as the minimal and self-consistent realization of the dust-interaction framework. In contrast to approaches invoking ground-state energy or quantum vacuum contributions, no a priori quantum interpretation is required for the constant $c_1$, and the notorious fine-tuning problem associated with the cosmological constant is entirely circumvented. Crucially, the term proportional to $c_1$ embodies the collective influence of local interactions among the constituent elements of the matter field, providing a natural and physically motivated mechanism that renders the coincidence problem inoperative. In this formulation, the emergent cosmological behavior arises entirely from the intrinsic dynamical structure of the matter sector, without the introduction of exotic fields, negative-pressure components, or additional degrees of freedom. The framework therefore not only reproduces the background evolution characteristic of $\Lambda$CDM but does so with a fundamentally distinct and conceptually transparent origin. This highlights both the theoretical elegance and the practical utility of the model, demonstrating that late-time cosmic acceleration can emerge intrinsically from the internal dynamics of conventional matter, while simultaneously resolving long-standing conceptual tensions in cosmology associated with fine-tuning and coincidence in $\Lambda$CDM model.

\section{Concluding Remarks and Implications}

In this work, we have revisited the foundational interpretation of the cosmological constant by focusing on the concept of ground-state energy in quantum field theory. We demonstrated that the standard identification of geometrical $\Lambda$ with vacuum energy introduces a structural inconsistency: while the quantum vacuum possesses energy without momentum flux, the conventional geometric implementation of $\Lambda$ enforces isotropic contributions to both temporal and spatial components of the Einstein field equations. This mismatch underscores the need for a framework in which the intrinsic temporal character of vacuum energy is consistently realized within gravitational dynamics.

By introducing the notion of energy--momentum symmetry breaking, we formulated a covariant mechanism whereby the temporal component of the modification sector contributes energy without an accompanying isotropic pressure. Applying this construction to a spatially flat Friedmann--Robertson--Walker spacetime revealed that the interaction between the matter sector and the symmetry-breaking vacuum naturally induces an effective pressure, driving late-time cosmic acceleration. Unlike phenomenological prescriptions, the interaction term $\mathcal{Q}$ emerges directly from the underlying structure of the field equations, establishing a dynamical coupling between matter and vacuum contributions. This approach resolves both the fine-tuning and coincidence problems of the $\Lambda$CDM paradigm: the proportionality constant connecting ground-state energy to its macroscopic manifestation is fixed by fundamental scales, while the interaction ensures that matter and vacuum energy evolve in a correlated manner.

Extending this analysis, we illustrated that even in the absence of an explicit dark energy sector, the collective dynamics of ordinary dust matter suffice to reproduce $\Lambda$CDM–like expansion. By distinguishing two dynamically identifiable sectors within the dust matter field and allowing intrinsic interactions between them, a macroscopic effective pressure naturally emerges. On sufficiently large scales, the matter field retains its dust-like character, yet locally the energy exchange between sectors generates a global contribution analogous to the cosmological constant. In the minimal symmetric realization, with $\alpha=\beta=0.5$ and $\gamma=-0.5$, the model reproduces the $\Lambda$CDM background evolution exactly, without invoking exotic components, quantum vacuum interpretations, or additional degrees of freedom. The term proportional to $c_1$ directly encodes the internal interactions among the constituents of the matter field, rendering the coincidence problem irrelevant and entirely bypassing the fine-tuning issue.

Markov Chain Monte Carlo analysis using the Pantheon+SH0ES dataset confirms that the dust-interaction model is quantitatively consistent with observational constraints. The best-fit parameter set demonstrates remarkable agreement with the standard cosmological constant scenario. The corresponding evolution of the total equation-of-state parameter $\omega_\mathrm{tot}$ and the deceleration parameter $q$ closely tracks $\Lambda$CDM predictions, highlighting the viability of a dust-only, symmetry-breaking realization of late-time acceleration.

In summary, our investigation establishes that the $\Lambda$CDM model can be obtained either via a ground-state energy formalism or, more parsimoniously, as the emergent outcome of intrinsic interactions within ordinary dust matter. The latter approach achieves the $\Lambda$CDM expansion history without invoking additional fields or quantum interpretations, while simultaneously circumventing the fine-tuning and coincidence problems. Conceptually, this framework provides a transparent and physically motivated origin for cosmic acceleration: it emerges dynamically from the internal structure of matter itself. The results highlight the potential of energy--momentum symmetry breaking as a unifying principle in cosmology and open a pathway for further explorations of large-scale structure dynamics, cosmological perturbations, and the origin of late-time acceleration within standard General Relativity and even alternative theories in gravity.
\par\vspace{1em}
\noindent

\acknowledgments{HF thanks the Research Council of UFES for financial support. HF also thanks A.H. Fazlollahi for suggesting the consideration of this model.}
\par\vspace{1em}
\noindent
Data Availability Statement: No Data associated in the manuscript

\end{document}